\documentclass[smallextended,numbook,runningheads]{svjour3}     
\smartqed  
\usepackage{graphicx,amsmath,amssymb,bm,color,geometry}
\usepackage{eucal}
\usepackage[pagewise,displaymath]{lineno}
\usepackage{algorithmic,algorithm,comment}
\usepackage{tabularx}
\usepackage{booktabs,cleveref,tikz}

\journalname{}

\usepackage{subcaption}
\DeclareCaptionLabelFormat{subcaptionlabel}{\normalfont(\textbf{#2}\normalfont)}
\begin{document}

\title{Estimating Failure Probability with Neural Operator Hybrid Approach}


\author{Mujing Li \and Yani Feng \and Guanjie Wang
}


\institute{
Mujing Li \at School of Information Science and Technology, ShanghaiTech University, Shanghai, China.\\\ \email{limj@shanghaitech.edu.cn} \\\
Yani Feng \at School of Information Science and Technology, ShanghaiTech University, Shanghai, China.\\\ \email{fengyn@shanghaitech.edu.cn} \\\
Guanjie Wang  \at School of Statistics and Mathematics, Shanghai Lixin University of Accounting and Finance,
Shanghai, China\\\ \email{guanjie@lixin.edu.cn (corresponding author)}  \\\ 
}

\date{Received: date / Accepted: date}
\maketitle

\begin{abstract}
Evaluating failure probability for complex engineering systems is a computationally intensive task. While the Monte Carlo method is easy to implement, it converges slowly and, hence, requires numerous repeated simulations of a complex system to generate sufficient samples. To improve the efficiency, methods based on surrogate models are proposed  to approximate the limit state function. In this work, we reframe the approximation of the limit state function as an operator learning problem and utilize the DeepONet framework with a hybrid approach to estimate the failure probability. The numerical results show that our proposed method outperforms the prior neural hybrid method.

\end{abstract}

\section{Introduction}

{In practice, evaluating failure probability for systems that inherently contain uncertainty is a fundamental problem encountered in various fields, such as structural safety, risk management,~reliability-based optimization, etc. Uncertainties in such systems are abstracted in terms of the failure mode, and~failure probability estimation is essentially a problem of evaluating multivariate integrals in domains defined by certain failure modes. While the mathematical formulation of the problem is well defined, evaluating  such integrals remains a challenging task in practice.}

The most straightforward approach to evaluating the failure probability is to use the Monte Carlo sampling (MCS) method~\cite{XiuMC3,XiuMC8}. However, due to its slow convergence, MCS requires numerous samples, resulting in a heavy computational burden. This computational burden becomes even more pronounced when complex stochastic PDEs represent the failure modes, since MCS necessitates repeatedly solving the model to estimate the failure~probability.

To address this issue, various approaches have been developed, including the first-order reliability method (FORM) \cite{like3FORM}, second-order reliability method (SORM) \cite{like4SORM}, and~response surface method (RSM) \cite{like5RSM,like14}. These methods replace the limit state function with a surrogate model that is easy to evaluate, thereby greatly reducing the simulation time. Following this idea, a~hybrid method was proposed by~\cite{Xiu2010,Xiu2011} to estimate the probability based on the surrogate model while re-evaluating samples in a given suspicious region. The~design of the hybrid method significantly reduces the time complexity while ensuring~accuracy.

There are various methodologies for constructing surrogate models, such as the stochastic Galerkin method~\cite{Ghanem2003,Xiu2002wiener}, the~reduced basis method~\cite{Boyaval2010Reduced,Quarteroni2016Reduced}, and~deep learning. Deep learning has rapidly developed in recent decades, particularly in scientific and engineering applications. Physics-informed neural networks (PINNs), which build upon the widely known universal approximation capabilities of continuous functions for neural networks (NNs) \cite{cybenko1989approximation,hornik1989multilayer}, were introduced in~\cite{2019PINN} and have demonstrated their efficiency in numerous studies~\cite{lu2021deepxde,pang2019fpinns,zhang2020learning}. By~utilizing established deep learning and machine learning techniques, NN models can be employed as surrogate models to approximate the limit state function, outperforming traditional surrogate models in certain problems, such as those of high-dimensional systems~\cite{Li2019,Lieu2022,mei2022}.

{As a significant area within the domain of deep learning, operator learning has emerged in recent years. The~underlying principle of operator learning resides in the observation that nonlinear operators can be effectively approximated by employing single-layer neural networks}
\cite{chen1993approximations,chen1995approximation} (\Cref{thm:1}). {Operator learning} aims to map infinite-dimensional functions to infinite-dimensional functions. Since it is more expressive and can break the curse of dimensionality in input space~\cite{Lu2022},~operator learning has gained much attention in recent years~\cite{Lu2022,FNO2020,Lu2021}. Among~the operator learning techniques, the~DeepONet introduced in~\cite{Lu2021,Lu2022} has been demonstrated to be effective in numerous applications, including~\cite{DON40,DON41,DON43}.


In this work, we present a novel approach to  failure probability estimation by reframing the approximation problem of the limit state function as an operator learning problem and~subsequently adapting the DeepONet framework to address it. The~operator learning formulation provides a more effective and generalized approach to constructing a surrogate model for the limit state function, resulting in enhanced precision and reduced simulation numbers. To~further ensure the precision, we employed a hybrid method~\cite{Xiu2010} for estimating the failure probability. Our proposed neural operator hybrid (NOH) approach significantly reduces the time complexity while maintaining high accuracy compared to earlier neural hybrid and Monte Carlo simulation approaches. We posit that the efficiency of our approach in estimating failure probability demonstrates the potential of operator learning in various~tasks.

This paper is structured as follows: In \Cref{sec:bg}, we present the problem setting and introduce a hybrid method for evaluating the failure probability. The~neural operator learning and proposed algorithm are then fully described in \Cref{sec:met}. To~demonstrate the effectiveness of our approach, we describe numerical experiments in \Cref{sec:exp} that cover a variety of scenarios, including ODEs, PDEs, and~multivariate models. Finally, we offer concluding remarks and observations in  \Cref{sec:conclusions}.

\section{Preliminaries}
\label{sec:bg}
This section will provide an overview of the mathematical framework for failure probability and introduce a hybrid method for solving this~problem.
\subsection{Problem~Setting}\label{sec:pb}
Let $Z=\left(Z_1, Z_2, \ldots, Z_{n_z}\right)$ be an $n_z$-dimensional random vector with the distribution function $F_Z(z)=\operatorname{Prob}(Z \leq z)$. The~image of $Z$, i.e.,~the set of all possible values that $Z$ can take, is denoted by $\Omega$. It is our interest to evaluate the failure probability $P_f$ defined by
\Cref{eq:fp}:
\begin{linenomath}
	\begin{equation}\label{eq:fp}
	P_f=\operatorname{Prob}\left(Z \in \Omega_f\right)=\int_{\Omega_f} \mathrm{d} F_Z(z)=\int \chi_{\Omega_f}(z) \mathrm{d} F_Z(z)= \mathbb{E}[\chi_{\Omega_f}(z)],
	\end{equation}
\end{linenomath}
where the characteristic function $\chi_{\Omega_f}(z)$ is defined as:
\begin{linenomath}
	\begin{equation}\label{eq:characteristic}
	\chi_{\Omega_f}(z)=\left\{\begin{array}{lll}
	1 & \text { if } & z \in \Omega_f, \\
	0 & \text { if } & z \notin \Omega_f,
	\end{array}\right.
	\end{equation}
\end{linenomath} 
and the failure domain $\Omega_f$, where failure occurs, is defined as: 
\begin{linenomath}
	\begin{equation}\label{eq:fd}
	\Omega_f = \{Z: g(Z)<0\}.
	\end{equation}
\end{linenomath} 

Here, $g(Z)$ is a scalar limit state function---also known as a performance function---that characterizes the failure domain. It should be emphasized that, in~many real-world systems, $g(Z)$ does not have an analytical expression and is instead characterized by a complex system that requires expensive simulations to evaluate. Consequently, the~evaluation of $g(Z)$ can be computationally expensive, leading to significant time~complexity.

\subsection{Hybrid~Method}
\label{sec:hybrid}
The most straightforward approach to estimating  failure probability is the Monte Carlo sampling (MCS) method~\cite{XiuMC3,XiuMC8}, which is given by:
\begin{linenomath}
	\begin{equation}\label{pmc}
	P_f^{m c}=\frac{1}{M} \sum_{i=1}^M \chi_{\{g(z)<0\}}\left(z^{(i)}\right),
	\end{equation}
\end{linenomath}
where $\{z^{(i)}\}_{i=1}^M$ is a set of sample points for the random vector $Z$. The~characteristic function $\chi_{\{{g(z)<0}\}}(z^{(i)})$ takes a value of $1$ if the limit state function $g(Z)$ evaluated at $z^{(i)}$ is less than zero and~$0$ otherwise. The~failure probability $P_f^{m c}$ is estimated as the average of the characteristic function over the $M$ sample~points.

However, evaluating the limit state function $g(Z)$ at numerous sample points can be a computationally intensive task, especially when dealing with complex stochastic systems, resulting in significant simulation time complexity. To~address this issue, a~surrogate model can be used to approximate the limit state function $g(Z)$ and avoid the need for direct evaluation at each sample point. Specifically, a~surrogate model of $g(Z)$ is denoted by $\hat{g}(Z)$, which can be rapidly evaluated. The~failure probability can then be estimated as:
\begin{linenomath}
	\begin{equation}
	\hat{P}_f^{m c}=\frac{1}{M} \sum_{i=1}^M \chi_{\{\hat{g}(z)<0\}}\left(z^{(i)}\right).
	\end{equation}
\end{linenomath}

While surrogate models can significantly reduce computational costs in Monte Carlo methods, relying solely on them for estimating the failure probability may result in poor precision or even failure. To~address this issue,  a~hybrid approach that combines the surrogate models $\hat{g}$ and the limit state function $g$ was proposed in~\cite{Xiu2010,Xiu2011}. In~the following, we give a brief review of the hybrid~method.

Suppose that $(-\gamma, \gamma)$ is a suspicious region, where $\gamma$ is a non-negative real number. In~this case, we can approximate the failure domain $\Omega_f$ with  $\widetilde{\Omega}_f$ as follows:        
\begin{linenomath}
	\begin{equation}
	\widetilde{\Omega}_f = \left\{\hat{g}(Z)<-\gamma\right\} \cup\left\{\left\{\left|\hat{g}(Z)\right| \leq \gamma\right\} \cap\{g(Z)<0\}\right\},
	\end{equation}
\end{linenomath}
where $g$ is the limit state function, and~$\hat{g}$ represents the surrogate model of $g$.  Enhanced with the hybrid method, the~failure probability can be estimable by MCS:
\begin{linenomath}
	\begin{equation}
	\begin{aligned}
	P_f^h &=\frac{1}{M} \sum_{i=1}^M \chi_{\hat{\Omega}_f}\left(z^{(i)}\right)\\
	&=\frac{1}{M} \sum_{i=1}^M\left[\chi_{\{\hat{g}<-\gamma\}}\left(z^{(i)}\right)+\chi_{\{|\hat{g}| \leq \gamma\}}\left(z^{(i)}\right) \cdot \chi_{\{g<0\}}\left(z^{(i)}\right)\right].
	\end{aligned}
	\end{equation}
\end{linenomath}

The hybrid method can be considered as an approach for estimating $P_f$ by using a surrogate $\hat{g}$, followed by a re-evaluation of the samples within the suspicious domain. While increasing the value of $\gamma$ leads to higher time complexity, it also results in more accurate estimation. In~Ref.~\cite{Xiu2010}, it was proved that for any surrogate $\hat{g}(Z)$ and for all $\epsilon>0$, there exists a critical value $\gamma_N>0$ such that for all $\gamma>\gamma_N$, the~difference between the estimated $P_f^h$ and the truth $P_f$ is less than $\epsilon$, i.e.,
\begin{linenomath}
	\begin{equation}
	\left|P_f-P_f^h\right|<\epsilon.
	\end{equation}
\end{linenomath}

To be more precise,
\begin{linenomath}
	\begin{equation}
	\gamma_N=\frac{1}{\epsilon^{1 / p}}\left\|g(Z)-\hat{g}(Z)\right\|_{L_{\Omega}^p},
	\end{equation}
\end{linenomath}
where the approximation is measured in the $L^p$-norm with $p \geq 1$.
\begin{linenomath}
	\begin{equation}
	\left\|g(Z)-\hat{g}(Z)\right\|_{L_{\Omega}^p}=\left(\int_{\Omega}\left|g(z)-\hat{g}(z)\right|^p \mathrm{d} F_Z(z)\right)^{1 / p}.
	\end{equation}
\end{linenomath}

Selecting an appropriate value of $\gamma$ that balances accuracy and computational efficiency can be a challenging task. To~address this challenge, an~iterative algorithm, as~demonstrated in \Cref{alg:h1}, is commonly employed in practice instead of directly selecting $\gamma$. In~\Cref{alg:h1},  the~surrogate $\hat{g}$ samples are gradually replaced with $g$ samples in the iteration procedure until either the stopping criterion is reached or~the iteration step reaches $\lceil M / \delta M\rceil$, which is equivalent to expanding the suspicious region at each iteration. When $k$ reaches $\lceil M / \delta M\rceil$, the~iterative hybrid algorithm degenerates to the Monte Carlo method~\eqref{pmc}, indicating that the convergence is achieved as $P_f^{(k)} \rightarrow P_f^{m c}, \quad k \rightarrow\lceil M / \delta M\rceil$. 
It is obvious that the time complexity of the iterative hybrid algorithm is heavily influenced by the accuracy of the surrogate model used in \Cref{alg:h1}.

\begin{algorithm}
	\caption{Iterative Hybrid Method~\cite{Xiu2010}.}
	\label{alg:h1}
	\begin{algorithmic}[1]
		\STATE \textbf{Input:} surrogate model $\hat{g}$, $ S= \{z^{(i)}\}_{i=1}^M$ samples from random variable $Z$, tolerance $\epsilon $, and sample size in each iteration $\delta M$.
		\STATE Initialization: $ k=0 $.
		\STATE  Compute $P_f^{(0)}=\frac{1}{M} \sum_{i=1}^M \chi_{\{\hat{g}(z)<0\}}\left(z^{(i)}\right)$.
		\STATE Sort $\left\{\left|\hat{g}\left(z^{(i)}\right)\right|\right\}_{i=1}^M$ in ascending order; sort the correspond sample $S$ accordingly.
		\FOR{ $k$ from 1 to $ \lceil M / \delta M\rceil$ }
		\STATE $ \delta S^k = \left\{z^{(j)}\right\}_{j=(k-1) \delta M+1}^{k \delta M}$.
		\STATE $\delta P=\frac{1}{M} \sum_{z^{(j)} \in \delta S^{k}}\left[-\chi_{\{\hat{g}<0\}}\left(z^{(j)}\right)+\chi_{\{g<0\}}\left(z^{(j)}\right)\right]$.
		\STATE $P_f^{(k)} = P_f^{(k-1)} + \delta P$.
		\IF {$|\delta P| \leq \epsilon$ for several times}
		\STATE \textbf{break}
		\ENDIF
		\ENDFOR
		\STATE \textbf{Output:} $P_f^{(k)}$
	\end{algorithmic}
\end{algorithm}

\section{Neural Operator Hybrid~Algorithm}\label{sec:met}
In \Cref{sec:bg}, we described the failure probability problem and the hybrid algorithm for solving it. As~we have mentioned, the~accuracy of the surrogate model greatly affects the performance of the iterative hybrid algorithm. In~this section, we introduce  neural operator learning and present the neural operator hybrid (NOH) algorithm, which reframes the approximation problem of the limit state function as an operator learning problem. Unlike prior studies that used neural networks as surrogate models for mappings \mbox{(as in~\cite{Li2019,Lieu2022,mei2022,papadrakakis2002reliability,kutylowska2015neural}),} our algorithm constructs a surrogate model by using operator learning techniques. {The benefits of our method are mainly in two aspects: First, it increases the generalization of the surrogate model. Second, it increases the precision of the surrogate model with more information involved, which} results in a more effective and generalized approach to estimating failure~probability.
\subsection{Neural Operator~Learning}
Neural operator learning aims to accurately represent linear and nonlinear operators that map input functions into output functions. More specifically, let $U$ be a vector space of functions on set $K_1$, and~let $V$ be a vector space of functions on set $K_2$; $G$ is an operator map from $U$ to $V$, i.e.,
\begin{linenomath}
	\begin{equation} 
	G : u  \mapsto G(u) \in V,\text{ for } u \in U,   
	\end{equation}
\end{linenomath}
where $u$ is a function defined on the domain $K_1$, i.e.,
\begin{linenomath}
	\begin{equation} 
	u : x \mapsto u(x) \in \mathbb{R}, \text{ for } x \in K_1,  
	\end{equation}
\end{linenomath}
and $G(u)$ is a function defined on the domain $K_2$, i.e.,
\begin{linenomath}
	\begin{equation} 
	G(u) : y \mapsto G(u)(y) \in \mathbb{R}, \text{ for } y \in K_2. 
	\end{equation}
\end{linenomath}

In the context of this paper, $U$ is referred to as the input function space, and~$V$ is referred to as the output function space. It is of interest to design neural networks that can approximate the mapping of the operator $G$  from the input function space to the output function~space.

In this work, we employ the DeepONet framework~\cite{Lu2021}, an~ascending 
operator learning approach based on the following theorem, to~construct the surrogate of the operator $G$. 
\begin{theorem}[Universal Approximation Theorem for the Operator~\cite{chen1995approximation}]
	\label{thm:1} 
	Suppose that $\sigma$ is a continuous non-polynomial function, $X$ is a Banach Space, $K_1 \subset X, K_2 \subset \mathbb{R}^d$ are two compact sets in $X$ and $\mathbb{R}^d$, respectively, $U$ is a compact set in $C\left(K_1\right)$, and $G$ is a nonlinear continuous operator that maps $U$ into $C\left(K_2\right)$. Then, for any $\epsilon>0$, there are positive integers $n, p, m$ and constants $c_i^k, \xi_{i j}^k, \theta_i^k, \zeta_k \in \mathbb{R}, w_k \in \mathbb{R}^d, x_j \in K_1$, $i=1, \ldots, n, k=1, \ldots, p, j=1, \ldots, m$, such that
	\begin{linenomath}
		\begin{equation} 
		|G(u)(y)-\sum_{k=1}^p \underbrace{\sum_{i=1}^n c_i^k \sigma\left(\sum_{j=1}^m \xi_{i j}^k u\left(x_j\right)+\theta_i^k\right)}_{\text {branch }} \underbrace{\sigma\left(w_k \cdot y+\zeta_k\right)}_{\text {trunk }}|<\epsilon.
		\end{equation}
	\end{linenomath}
	holds for all $u \in U$ and $y \in K_2$.
\end{theorem}

\begin{figure}[!htbp]
	\hspace{-18pt}
	\includegraphics[width=0.98\textwidth]{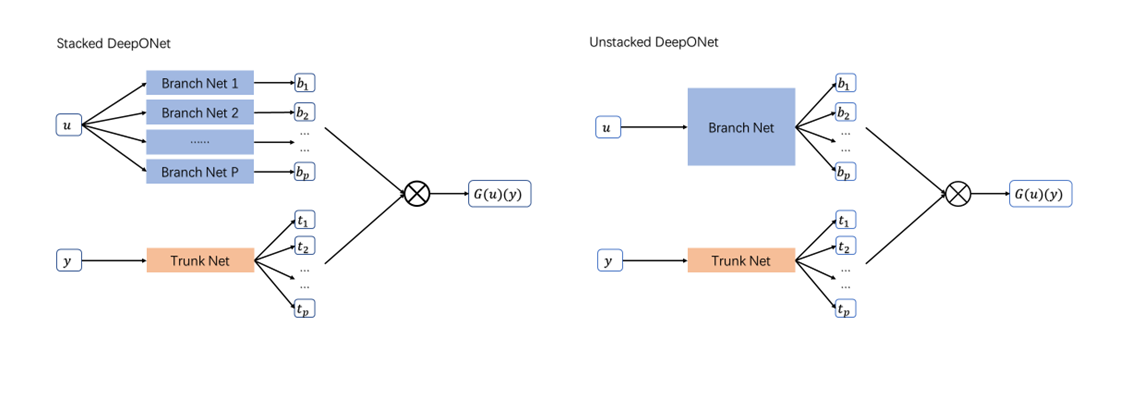}
	\vspace{-12pt}
	\caption{Illustrations of stacked and unstacked architectures of DeepONets~\cite{Lu2021}. DeepONets with unstacked structures  replace branch networks with a single multi-layer neural~network.}\label{fig:DeepONet}
\end{figure}

In DeepONet, the~operator $G$ is approximated by taking the inner product of two components, which can be expressed as follows:
\begin{linenomath}
	\begin{equation} 
	G(u)(y) \approx \mathcal{G}(u)(y):=\sum_{k=1}^p \underbrace{b_k(u)}_{\text {branch }} \underbrace{t_k(y)}_{\text {trunk }},
	\end{equation}
\end{linenomath}
where $b_k(u)$ is the output of the \textbf{{trunk network}} for a given input function $u$ in $U$, and~$t_k(y)$ is the output of the \textbf{branch network} for a given $y$ in $K_2$. \Cref{fig:DeepONet} illustrates this~architecture. 

In practical applications, the~input function $u$ is vectorized as $\left[u\left(x_1\right), \ldots, u\left(x_m\right)\right]^T$, where the sample points $[x_1,x_2 \ldots x_m]^T$ are referred to as \textbf{{sensors}}. The~input of the trunk network takes $P$ specific values $y_1^{(i)},\ldots,y_P^{(i)}$ in $K_2$ for each $G(u^{(i)})$. The~neural network can be trained by using the training data:
\begin{linenomath}
	\begin{equation}\label{generate_data}
	\mathcal{T}=\left\{\left(u^{(1)}, G(u^{(1)})\right),\left(u^{(2)}, G(u^{(2)})\right), \cdots,\left(u^{(N)}, G(u^{(N)})\right)\right\},
	\end{equation}
\end{linenomath}
and by minimizing the following loss function:
\begin{linenomath}
	\begin{equation}
	\mathcal{L}(\Theta)=\frac{1}{N  P}\sum_{l=1}^{P}\sum_{i=1}^{N}{\Arrowvert G_{\Theta}(u^{(i)})(y_l^{(i)})-G(u^{(i)})(y_l^{(i)})\Arrowvert}_2 ,
	\end{equation}
\end{linenomath}
where the neural network $G_{\Theta}$ with parameter $\Theta$ approximates the operator $G$.

\subsection{Neural Operator Hybrid~Algorithm}
Neural operator learning involves functions as both inputs and outputs, requiring us to reframe our problem accordingly. Let $U$ be the collection of all functions defined on $[0,1]$ such that
\begin{linenomath}
	\begin{equation} 
	u(x) = \overline{Z} -k\overline{\sigma} + 2k\overline{\sigma} x, \text{ for } x\in [0,1],
	\end{equation}
\end{linenomath}
where $k$ is a selective parameter, and $\overline{Z}$ and $\overline{\sigma}$ are defined as:
\begin{linenomath}
	\begin{equation} 
	\overline{Z} = \frac{Z_1+\ldots+Z_{n_z}}{n_z},\ \overline{\sigma} = \frac{\sigma_1+\ldots+\sigma_{n_z}}{n_z}.
	\end{equation}
\end{linenomath}

Here, $\sigma_1,\ldots,\sigma_{n_z}$ are the variances of random~variables  
$Z_1,\ldots,Z_{n_z}$, respectively.  We can then define an operator $G$ from $U$ to $V$ as follows:
\begin{linenomath}
	\begin{equation} 
	G: u \mapsto G(u) \in V, \text{ for } u\in U, 
	\end{equation}
\end{linenomath}
where $V = \mathrm{span}\{g(x)\}$, and $G(u)$ is defined by
\begin{linenomath}
	\begin{equation}\label{eq:OLP_def}
	G(u): y \mapsto g(y) \in \mathbb{R}, \text{ for } y \in \mathbb{R}^{n_z}.
	\end{equation}
\end{linenomath}

Here, $g$ is the limit state function discussed in \Cref{sec:pb}. Then, designing a surrogate model for $G$ is a standard operator learning~problem.

The inspiration for this reframing is the establishment of a relationship between the input function $u(x)$ and the random variable $Z$.  It is important to note that the prior distribution of $u(x)$ is entirely known, which enables us to generate training data by using the following process: Firstly, we randomly sample $z^{(i)}$  according to the random distribution $Z$, where $i=1,\ldots,N$. Next, we define $u^{(i)}(x)$ by using the following equation:
\begin{linenomath}
	\begin{equation}\label{eq:uxdef}
	u^{(i)}(x) = \overline{z}^{(i)} -k\overline{\sigma} + 2k\overline{\sigma}x, \ x\in[0,1],
	\end{equation}
\end{linenomath}
where
\begin{linenomath}
	\begin{equation} 
	\overline{z}^{(i)} = \frac{z^{(i)}_1+\ldots+z^{(i)}_{n_z}}{n_z}.
	\end{equation}
\end{linenomath}

{In practical implementation, we vectorize the input function $u^{(i)}(x)$ on a uniform grid of the interval $[0,1]$. Specifically, the~vectorized input function is given by $[u^{(i)}(x_1),\ldots,u^{(i)}(x_m)]$,} where $x_j$ denotes the $j$-th sensor and is defined by
\begin{linenomath}
	\begin{equation} 
	x_j=\frac{j-1}{m-1}, \ j=1,\ldots,m.
	\end{equation}
\end{linenomath}

Suppose that there are $P$ observations $y_l^{(i)}\in \mathbb{R}^{n_z}$, $l=1,\ldots,P$ for $G(u^{(i)})$; then, by \mbox{Equation~\eqref{eq:OLP_def},} we have $G(u^{(i)})(y_l^{(i)})=g(y_l^{(i)})$. Once the dataset is generated, the~model is trained by minimizing the following loss function:
\begin{linenomath}
	\begin{equation} 
	\mathcal{L}(\Theta)=\frac{1}{N  P}\sum_{l=1}^{P}\sum_{i=1}^{N}{\Arrowvert G_{\Theta}(u^{(i)})(y_l^{(i)})-g(y_l^{(i)})\Arrowvert}_2,
	\end{equation}
\end{linenomath}
where $G_{\Theta}$ is a neural network with parameters $\Theta$ that approximates the operator $G$. 

\begin{figure}[!htbp]
	\begin{center}
		\begin{tikzpicture}
		\draw[rounded corners,black] (0,0) rectangle (5,2.4);
		\draw[->,thick,black] (5,1.2)--(7,1.2);
		\draw [rounded corners,black] (7,0) rectangle (12,2.4);
		\draw (2.4,1.2) node [above, black] {\small Generating training dataset:};
		\draw (2.4,1.2) node [below, black] {\small $\mathcal{T}=\left\{\left(u^{(i)},G(u^{(i)})\right)\right\}_{i=1:N}$};
		
		\draw (9.4,1.2) node [above, black] {\small Training operator learning};
		\draw (9.4,1.2) node [below, black] {\small neural network $G_{\Theta}$ using $\mathcal{T}$};
		\draw[rounded corners,black] (0,-4.4) rectangle (5,-2);
		\draw[->,thick,black] (7,-3.2)--(5,-3.2);
		\draw [rounded corners,black] (7,-4.4) rectangle (12,-2);
		\draw[->,thick,black]  (9.5,0) --(9.5,-2);
		
		\draw (9.4,-3.2) node [above, black] {\small Using $G_{\Theta}$ as the surrogate};
		\draw (9.4,-3.2) node [below, black] {\small model $\hat{g}$ in the hybrid Algorithm 1};
		
		\draw (2.45,-3.2) node [above, black] {\small {Estimating} the failure probability};
		\draw (2.4,-3.2) node [below, black] {\small $P_f$ using the hybrid algorithm};
		\end{tikzpicture}
	\end{center}
	\caption{A descriptive flowchart for the NOH~method.}
	\label{fig:alg}
\end{figure}
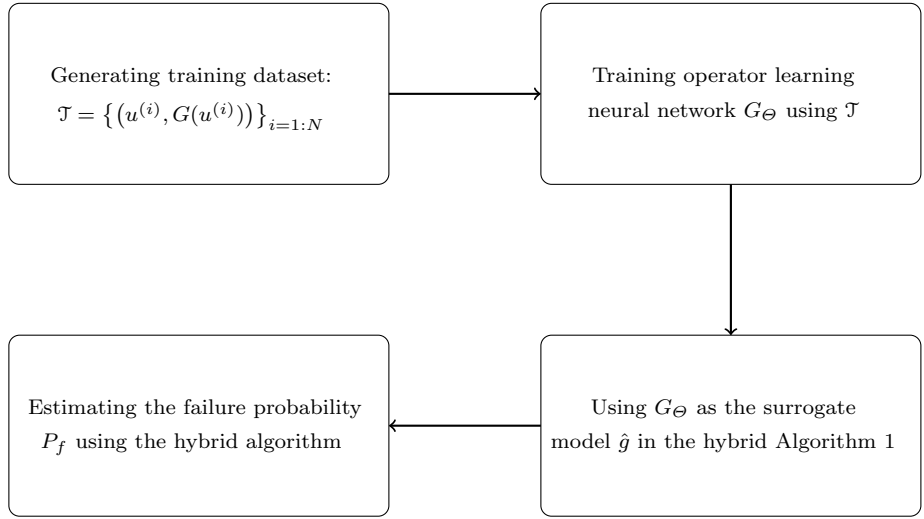

After constructing the surrogate for $G$ by using the neural network $G_{\Theta}$, we can integrate it into a hybrid algorithm to estimate the failure probability. This whole process is called the neural operator hybrid (NOH) method, and it is {shown in} \Cref{fig:alg}.

As discussed in Section~\ref{sec:hybrid}, the~convergence of Algorithm \ref{alg:h1} depends on the norm measurement of the difference between the surrogate and the limit state function. Therefore, it is crucial to have an accurate approximation $\hat{g}(Z)$ for reliable estimation. The~surrogate model constructed for the reformulated operator learning problem using DeepONet may achieve higher accuracy than that of the surrogate model constructed by using neural networks that do not incorporate information from random variables, as~the former model utilizes additional information from random distribution functions. Moreover, DeepONet exhibits less generalization error than that of simple neural networks~\cite{Lu2021}.

\unskip


\section{Numerical~Experiments}
\label{sec:exp}

In this section, we present three numerical examples to demonstrate the efficiency and effectiveness of the proposed neural operator hybrid (NOH) method. Furthermore, we compare the NOH method with the neural hybrid (NH) method.
For the purpose of clarity in presentation, we refer to the surrogate model constructed by using 
{fully connected} neural networks for $g$ in the NH method as \textbf{{the neural surrogate}} and~the surrogate constructed by using DeepONet for $g$ in the NOH method as \textbf{{the neural operator surrogate}}. {Both surrogate models are designed to approximate limit state function $g$, and~their main difference is the structure of neural networks utilized.}


For the NOH method, we use the simplest unstacked  DeepONet to construct the neural operator surrogate for the operator $G$, with~the branch and trunk networks implemented as fully connected neural networks (FNNs). The~trunk network is employed with a depth of 2 and a width of 40 FNNs, while the branch network has a depth of 2 and a width of 40 FNNs. To~facilitate a comparative analysis with the NH method, we built a neural surrogate for $g$ by using a simple FNN with a parameter size comparable to that of the NOH method. Specifically, in~the NH method, the~FNN utilized for the neural surrogate has a depth of 3, and~its width is adjusted to achieve a similar number of parameters to that in the DeepONet. Both models were optimized by using the Adam optimizer~\cite{Adam} with a learning rate of 0.001 on identical~datasets.

The code was run by using PyTorch~\cite{pytorch} and MATLAB 2019b on a workstation with an Nvidia GTX 1080Ti graphics card and~an Intel Core i5-7500 processor with 16 GB of RAM. It is noteworthy that the evaluation of time complexity is based on the performance function (PF) calls $N_{call}$, which refers to the number of system simulations that need to be executed, rather than the  running time of the programs, as~program running speeds may vary significantly across different programming languages and platforms. The~PF calls consist of the evaluation of the hybrid algorithm in line 6 of \Cref{alg:h1} and simulations for generating training data in~Equation~\eqref{generate_data}. 
We do not evaluate the computational time required for the neural surrogate or the neural operator surrogate, {as~a  model trained by using batch}  
techniques can evaluate $10^{5}$ samples in less than a~second.

\subsection{Ordinary Differential~Equation}
In this test problem, we consider a random ordinary differential equation (ODE) proposed in~\cite{Xiu2010}. The~ODE is given by:
\begin{linenomath}
	\begin{equation} 
	\frac{\mathrm{d} s}{\mathrm{d} t}=-Z s, \quad s(0)=s_0,
	\end{equation}
\end{linenomath}
where $s_0 = 1$, and~$Z \sim \mathcal{N}\left(\mu, \sigma^2\right)$ is a Gaussian random variable with a mean of $\mu = -2$ and standard deviation of $\sigma = 1$. The~limit state function is defined as $g(Z) = g(s(t, Z))=s(t, Z)-s_d$, where $s_d=0.5$ and $t = 1$. The~exact failure probability $P_f = 0.003539$ is regarded as the reference solution, which can be computed by using the analytic solution $s(t, Z)=s_0 \mathrm{e}^{-Z t}$.

To demonstrate the efficiency and effectiveness of the proposed NOH method, we compare it with a Monte Carlo simulation (MCS) and the NH method. We used DeepONet to train the neural operator surrogate in the NOH method and set the parameter $k$ in Equation~\eqref{eq:uxdef} to $4$, the~number of input functions for training $N$ to $500$, $P$ to $500$, and~the number of sensors $m$ to $100$. In~the NH method, we used the FNN as the neural~surrogate.

Both surrogates in the NH and NOH methods were trained with identical datasets, epochs, and~optimizers. Additionally, in~the MCS, $10^6$ samples were generated to estimate the failure~probability.

\Cref{tab:ex1} presents the performance of the MCS, the~NOH method, and the NH method. As~shown in the table, the~NOH method outperformed MCS by achieving the same level of estimation precision with only approximately $O(N_{call}/1000)$ or $0.23\%$ of the PF calls required by the MCS. The~NH method failed to estimate the failure probability, as~all of the outputs of the neural surrogate were greater than $0$. In~this special case, the~hybrid iterative procedure always terminated too early, while the estimated failure probability remained at $0$.

\begin{figure}[!htbp]
	    \centering 
	\includegraphics[width=0.8\textwidth]{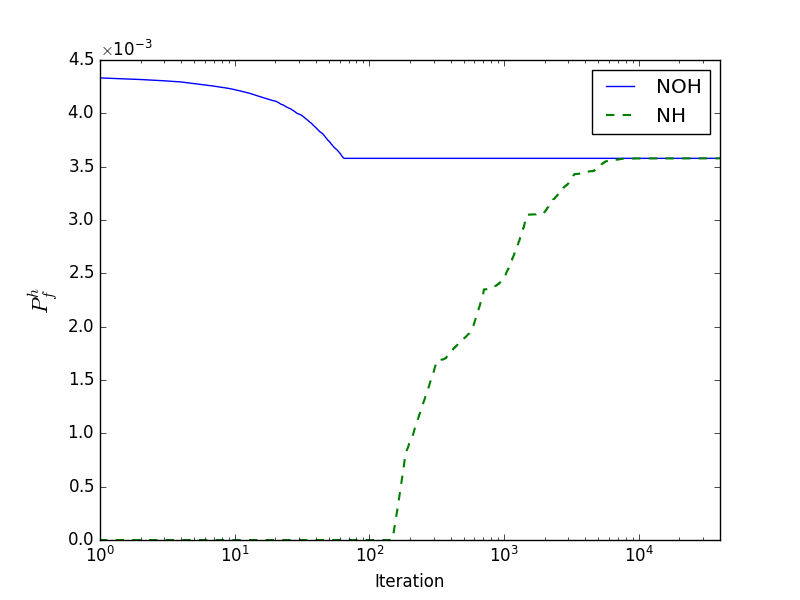} 
	\caption{Convergence of the NOH method and the NH method. In~the hybrid algorithm, we set $\delta M$ to $25$, and~the iterative procedure was not terminated until the limit state function $g$ was recomputed for $4\times10^4$ iterations.}
	\label{fig:Ex11}
\end{figure}

\begin{table}[!htbp]
	\caption{{Comparison} 
		of a Monte Carlo simulation (MCS), the~neural hybrid (NH) method, and the neural operator hybrid (NOH) method. In~the hybrid algorithm, we set $\delta M$ to $25$ and $\epsilon$ to $0$, and~we terminated the iterative procedure when $\delta P \leq \epsilon$ five times. The~relative error is denoted by $\epsilon_e$.}
	\label{tab:ex1} 
	\newcolumntype{C}{>{\centering\arraybackslash}X}
	\begin{tabularx}{\textwidth}{CCCC}
		\toprule
		\textbf{Method} & \boldmath $P_f^h$ & \boldmath $N_{call}$ & \boldmath $\epsilon_e$ \\ 
		\midrule
		MCS & $3.578 \times  10^{-3}$ & $10^6$ & $0.11\%$ \\  \midrule
		NOH & $3.578 \times  10^{-3}$ & 500 (Training) + 1750 (Evaluating) & 0.11\%  \\   \midrule
		NH & -  & - & - \\ 
		\bottomrule
	\end{tabularx} 
\end{table}
\Cref{fig:Ex11} illustrates the convergence of the NOH method and the NH method. In~order to compare the two methods, the~iterative procedure in the hybrid algorithm was not terminated until the limit state function $g$ was recomputed for at least $10^5$ samples. The~figure shows that the estimate of the failure probability by the NH method remained at $0$ until around $100$ iterations, and~it converged after approximately $7000$ iterations. In~contrast, the~NOH method converged after only $70$ iterations, demonstrating its superior efficiency compared to that of the NH~method.

\begin{figure}[!htbp]
	\centering 
	\begin{subfigure}[t]{0.49\textwidth}
		\centering 
		\includegraphics[width=1\textwidth]{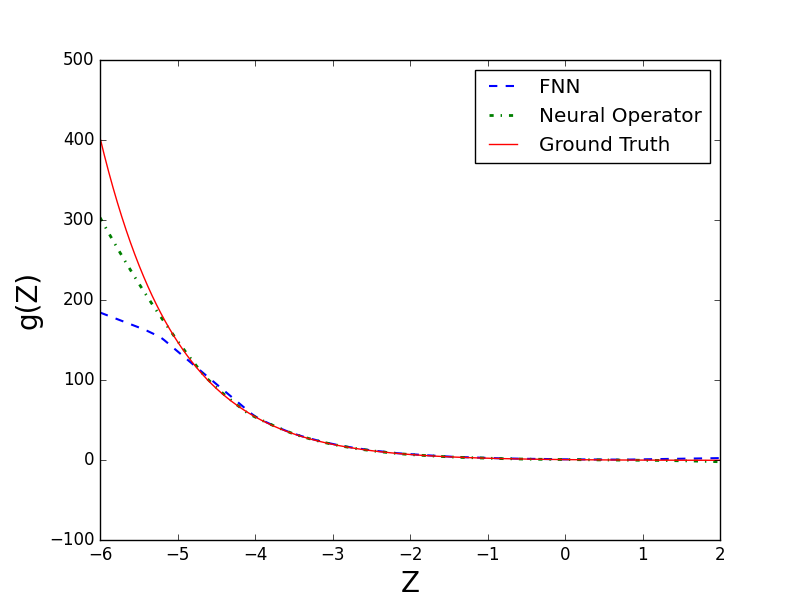}
		\caption{\centering  \label{fig:Ex12.a}}
	\end{subfigure}
	\hfill
	\begin{subfigure}[t]{0.49\textwidth}
		\includegraphics[width=1\textwidth]{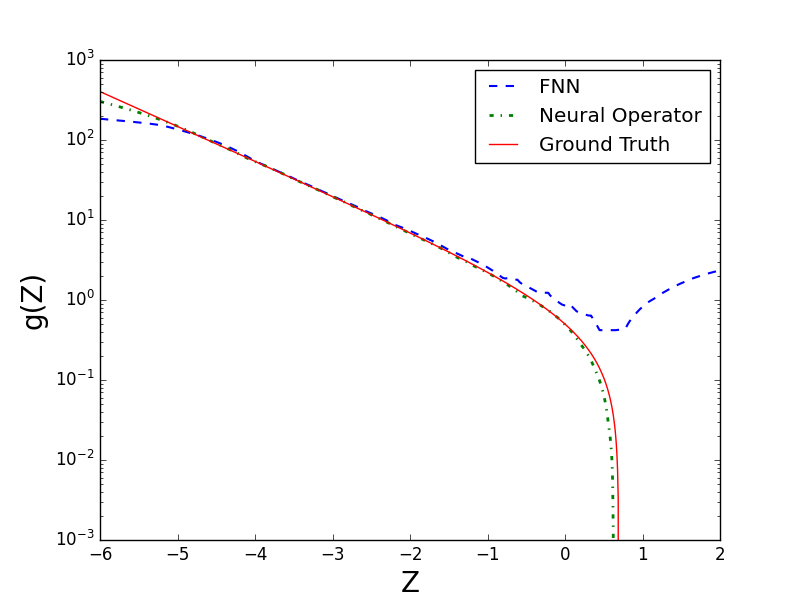}
		\caption{\centering  \label{fig:Ex12.b}}
	\end{subfigure}
	\caption{ Comparison of the neural surrogate and the neural operator surrogate. {Both models are compared to the ground truth $g(Z)$ from the analytic solution.} The right figure is the left figure on the log~scale. (\textbf{a}) Comparison of different surrogates. (\textbf{b}) Comparison on the  log {scale.} 
	} 
	\label{fig:Ex12}
\end{figure}

In \Cref{fig:Ex12}, we compare the performance of the neural surrogate and the neural operator surrogate in predicting the limit state function $g$. We observed that all of the outputs of the neural surrogate were greater than $0$, which led to the failure of the NH method in estimating the failure probability. It is evident that the neural operator surrogate outperformed the neural surrogate in predicting the limit state function $g$.

\subsection{Multivariate~Benchmark}
Next, we consider a high-dimensional multivariate benchmark problem (dimensionality: $n=50$) in the field of structural safety in~\cite{Li2019,Like25}:
\begin{linenomath}
	\begin{equation} 
	g(Z)=\beta n^{\frac{1}{2}}-\sum_{i=1}^n Z_i,
	\end{equation}
\end{linenomath}
where $\beta=3.5$ and each random variable $Z_i \sim \mathcal{N}(0,1),\,i=1,\dots,n$. $g(Z)$ is the limit state function. In~this test problem, the~reference failure probability is $P_f^{m c}=2.218 \times 10^{-4}$, which was obtained by using MCS with $5\times 10^6$ samples. 

The proposed NOH method is compared with the NH method in terms of accuracy and efficiency. For~the NOH method, we set the parameter $k$ in Equation~\eqref{eq:uxdef} to $4$, the~number of input functions $N$ to $1000$, $P$ to $1000$, and~the number of sensors $m$ to $100$. In~comparison, a~naive neural surrogate employing an FNN with a similar number of parameters was also constructed and trained under  conditions identical to those for the NH~method.

The performance of the MCS, the~NOH method, and~the NH method are illustrated in \Cref{tab:ex2}.
{Both the NOH method and the NH method demonstrated a substantial reduction in the number of samples required---approximately $0.1\%$ of the computational cost of MCS. Notably, the~NOH method outperformed the NH method by evaluating only $3\%$ of the $N_{call}$ while achieving a superior relative error $0.81\%$ compared to the NH method's relative error of $8.92\%$. This indicated that the NOH method achieved higher accuracy with significantly fewer samples, making it a more efficient and effective approach for the given task.}
\begin{table}[!htbp]
	\caption{A performance evaluation was conducted to compare the MCS, the~NH, and~the NOH approaches. The~NOH method was found to outperform the other methods, { and it only evaluated $3\%$ of  $N_{call}$ compared to NH method.} The NOH method also achieved the best relative error of $0.81\%$ while requiring the lowest number of samples with respect to the other~methods.}
	\footnotesize
	\label{tab:ex2}
	\newcolumntype{C}{>{\centering\arraybackslash}X}
	\begin{tabularx}{\textwidth}{CCCC}
		\toprule
		\textbf{Method} & \boldmath $P_f^h$ & \boldmath $N_{call}$ & \boldmath $ \epsilon_e$ \\ \midrule
		MCS & $2.22 \times 10^{-4}$ & $ 5 \times 10^6$ & - \\ \midrule
		NOH & $2.20 \times 10^{-4} $ & 1000 (Training)  + 150 (Evaluating) & 0.81\%  \\ \midrule
		NH &  $2.02 \times 10^{-4} $  & 1000 (Training)  + 4175 (Evaluating) & 8.92 \% \\ 
		\bottomrule
	\end{tabularx}
\end{table}

\begin{figure}[!htbp]
	    \centering 
	\begin{subfigure}[t]{0.49\textwidth}
		\includegraphics[width=1\textwidth]{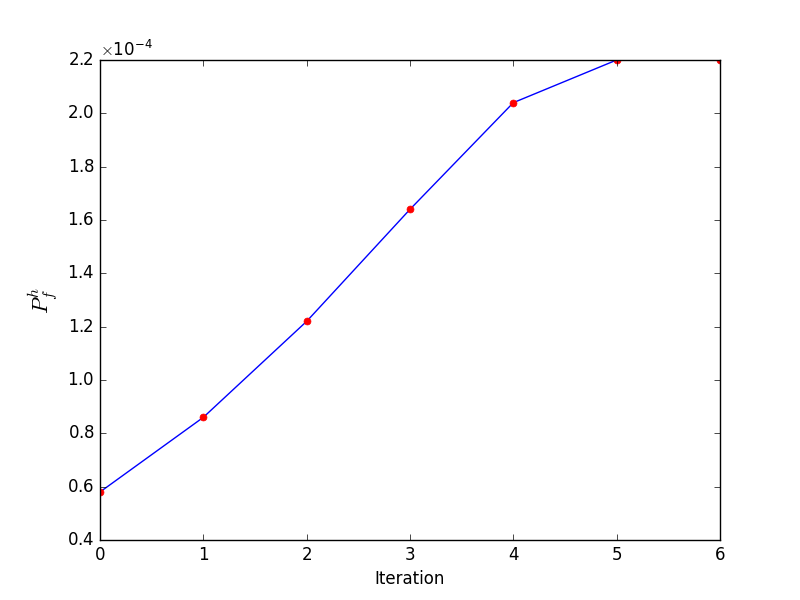}
		\caption{\centering  \label{fig:Ex22.a}}
	\end{subfigure}
	\begin{subfigure}[t]{0.49\textwidth}
		\includegraphics[width=1\textwidth]{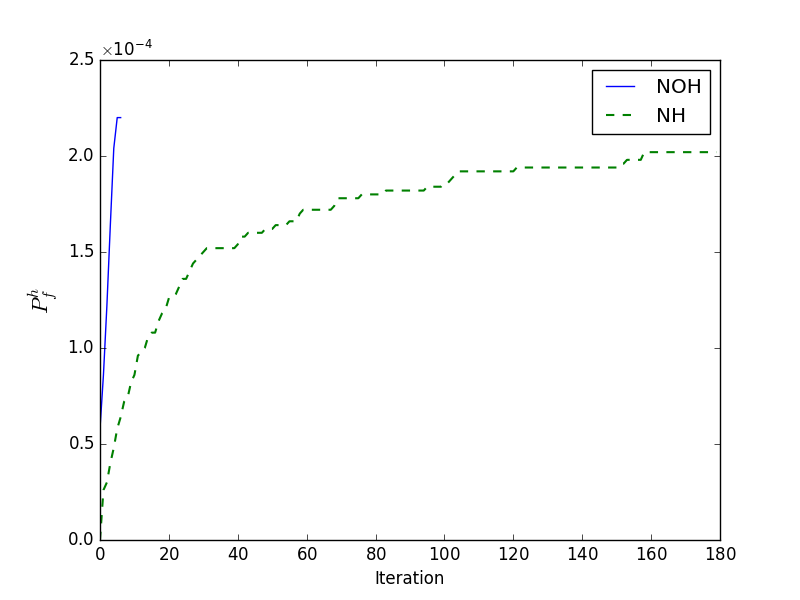}
		\caption{\centering  \label{fig:Ex22.b}}
	\end{subfigure}
	\vspace{-8pt}
	\caption{Convergence of the NOH method and the NH method. Panel~(\textbf{a}) demonstrates the convergence of the NOH method, where NOH terminates in six iterations, while in (\textbf{b}),  a comparison of the convergence rates between the NOH and the NH methods is conducted. {The NH method requires significantly more iterations than the NOH method to converge because its neural surrogate  lacks  precision.} For both methods, we set $\delta M = 25$ in each iteration. For~NOH, $\epsilon$ was set to $0$, and~the iterative procedure was terminated when $\delta P \leq \epsilon$ for the first~time. (\textbf{a}) Convergence of the NOH~method. (\textbf{b}) Comparison between the NOH and NH~{methods.} 
	}
	\label{fig:Ex22}
\end{figure}

In \Cref{fig:Ex22}, the~convergence of the NOH method is depicted and compared with that of the NH method. Figure~\ref{fig:Ex22}a demonstrates that the NOH method achieved an estimation of $P_f^h = 2.2\times10^{-4}$ with a relative error of $\epsilon_e = 0.81\%$ in less than six iterations. Figure~\ref{fig:Ex22}b 
{provides a comparison of convergent behaviors between the NH and NOH methods, clearly demonstrating that the NOH method converged in significantly fewer iterations. As~a consequence, the~NOH method required only $3\%$ of the total evaluations $N_{call}$. 
	These findings strongly suggest that neural operator surrogates offer improved precision and ease of training when compared to FNNs. 
	The reduced iteration times and lower number of evaluations highlight the superior efficiency and accuracy of neural operator surrogates in approximating complex functions or operators.}

\begin{figure}[!htbp]
    \centering

	\includegraphics[width=0.8\textwidth]{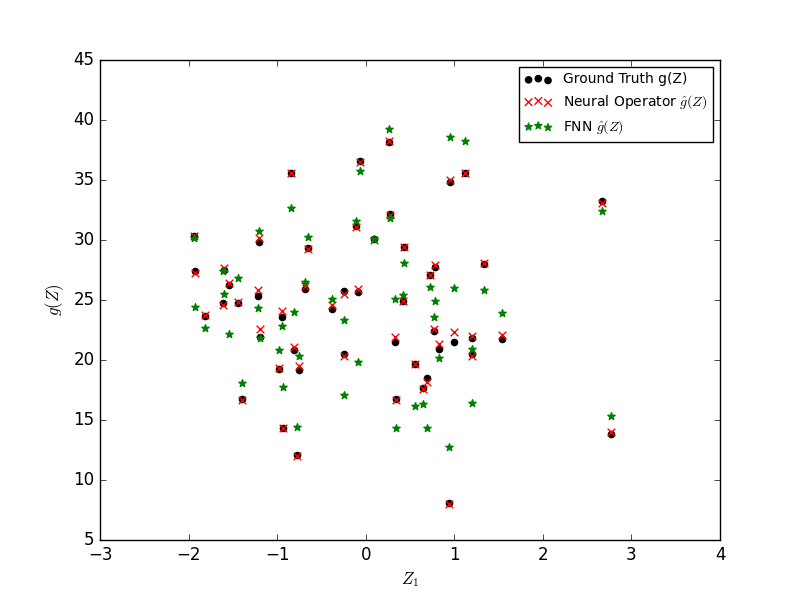}
	\caption{Comparison of the neural surrogate and the neural operator surrogate by using 50 randomly selected~samples.}
	\label{fig:Ex2.1}
\end{figure}

\Cref{fig:Ex2.1} 
{depicts the superior accuracy of the neural operator surrogates. It compares the neural surrogate and the neural operator surrogate approximations from $50$ randomly selected samples. As~illustrated in the figure, the~predictions of the limit function $g$ by the neural operator surrogate are more accurate than those of the neural surrogate, as~the former closely approximated the ground truth.}

{The experimental results provide evidence of the precision exhibited by the neural operator surrogate. The~combination of reduced iteration times and a lower number of evaluations further accentuates the efficiency and accuracy of neural operator surrogates in approximating complex functions or operators. These findings emphasize that the NOH method achieved higher levels of accuracy while utilizing significantly fewer samples.  }

\subsection{Helmholtz~Equation}
\begin{figure}[!htbp]
	   \centering 
	\includegraphics[width=1.0\textwidth]{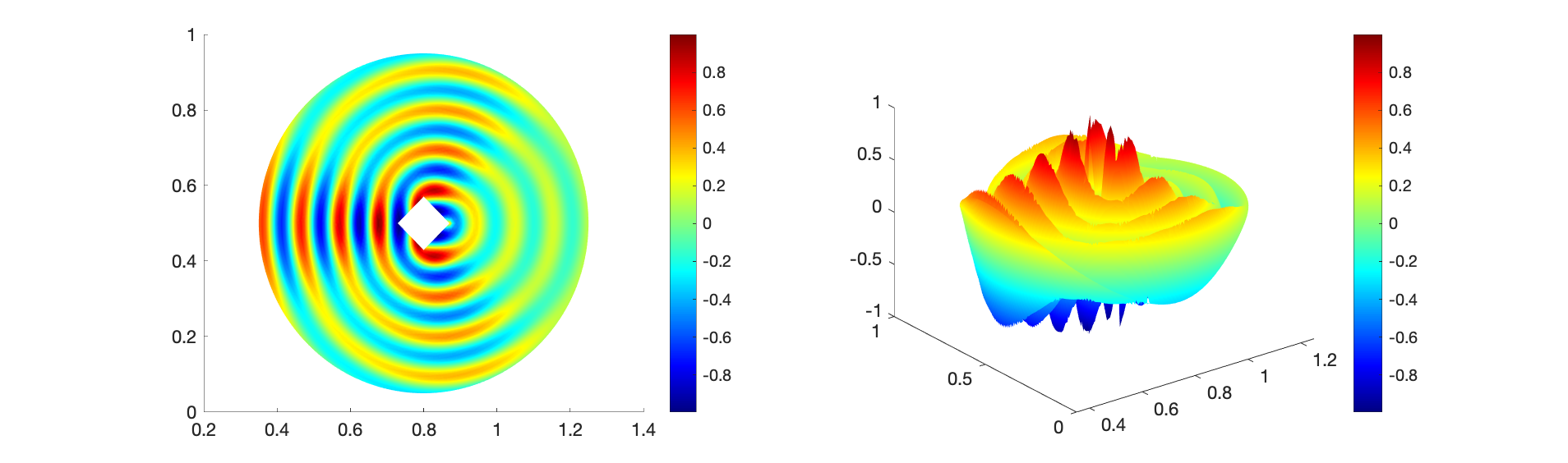} 
	\caption{{A snapshot} 
		of the solution of the Helmholtz~equation.}
	\label{fig:Ex3.1} 
\end{figure}

We consider the Helmholtz equation on a disk with a square hole~\cite{Li2019}. The~equation is given by:
\begin{linenomath}
	\begin{equation} 
	-\Delta u-\kappa^2 u=0,
	\end{equation}
\end{linenomath}
where coefficient $\kappa$ is a Gaussian random variable with a mean of $\mu = 60 $ and~variance of $\sigma=1$, i.e, $\kappa \sim \mathcal{N}\left(60, 1\right)$. The~system was set with a homogeneous term, and~Dirichlet boundary conditions ($u=0$) were applied on the edges of the square hole, while generalized Neumann conditions ($\vec{\xi} \cdot \nabla r-i \kappa r=0$, and $\vec{\xi}$ is the radial distance from the object) were applied on the edge of the disk. The~system was numerically solved by using the MATLAB PDE solver to obtain an accurate solution. A~snapshot of the solution of Helmholtz is shown in \Cref{fig:Ex3.1}. A~point sensor was placed at $x_p = [0.7264; 0.4912]$, and~the failure probability was defined as $\operatorname{Prob}(u(x_p, \kappa) > 1.00)$. The~reference solution was $P_f^{mc} = 2.70 \times 10^{-4}$, which was obtained with a  Monte Carlo simulation with $10^5$ samples.

With a setup similar to that in the previous experiment, we used DeepONet to train the neural operator surrogate in the NOH method and set the parameter $k$ in Equation~\eqref{eq:uxdef} to $4$, the~number of input functions $N$ to $1000$, $P$ to $1000$, and~the number of sensors $m$ to $100$. We used a fully connected neural network (FNN) as the neural surrogate in the NH method and~trained both models under identical~conditions.

In \Cref{tab:ex3}, 
{we present a performance analysis of the Monte Carlo simulation (MCS), the~NOH method, and~the NH method. Similarly to the previous experiments, the~results indicate that the NOH method required fewer $N_{call}$---about $11\%$---than the NH method did while providing a more accurate estimation with $3.70\%$ relative error versus $11.11\%$.}
\begin{table}[!htbp]
	\caption{A performance evaluation was conducted to compare the NH and the NOH approaches. The~NOH method was found to outperform the NH method {in both efficiency and accuracy}. Specifically, the~NOH method achieved a relative error of just $3.70\%$ while requiring fewer $N_{call}$ than the NH~method did. }
	\footnotesize
	\label{tab:ex3}
	\newcolumntype{C}{>{\centering\arraybackslash}X}
	\begin{tabularx}{\textwidth}{CCCC}
		\toprule
		\textbf{Method} & \boldmath $P_f^h$ & \boldmath $N_{call}$ & \boldmath $ \epsilon_e$ \\ \midrule
		MCS & $2.70 \times 10^{-4}$ & $10^5$ & - \\ \midrule
		NOH & $2.80 \times 10^{-4}$  & 1000 (Training)  + 100 (Evaluating) & 3.70\%  \\ \midrule
		NH & $3.00 \times 10^{-4}$  & 1000 (Training)  + 875 (Evaluating) & 11.11\% \\ 
		\bottomrule
	\end{tabularx}
\end{table}

\begin{figure}[!htbp]
	\centering 
	\begin{subfigure}[t]{0.49\textwidth}
		\centering 
		\includegraphics[width=1\textwidth]{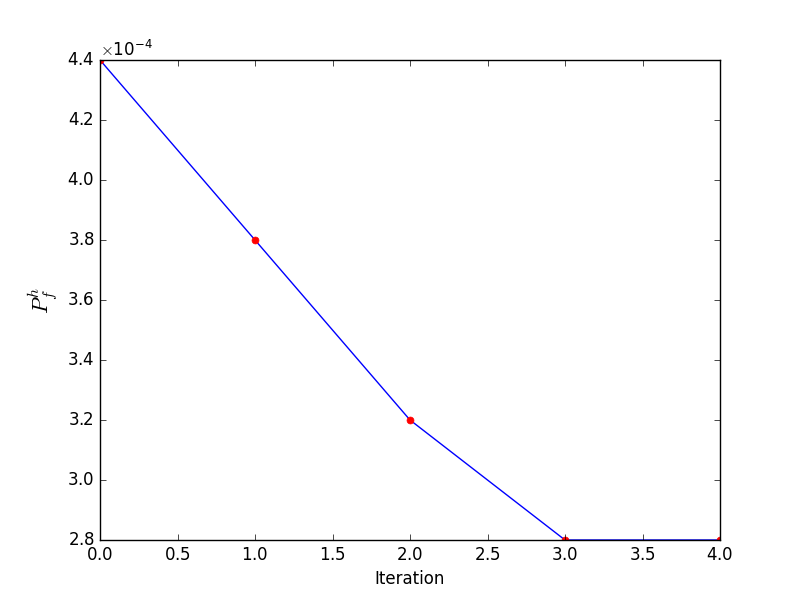}
		\caption{\centering \label{fig:Ex33.a}}
	\end{subfigure}
	\begin{subfigure}[t]{0.49\textwidth}
		\centering 
		\includegraphics[width=1\textwidth]{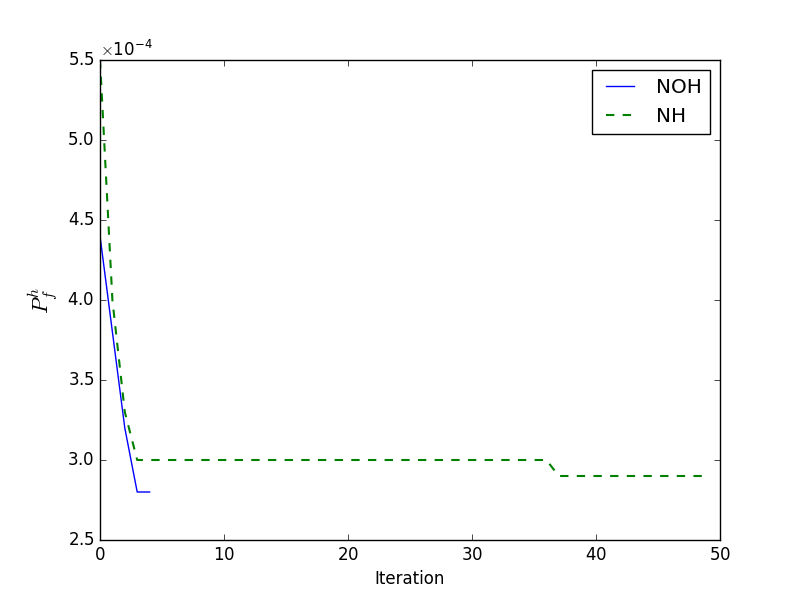}
		\caption{\centering \label{fig:Ex33.b}}
	\end{subfigure}
	\caption{{Convergence} 
		of the NOH method and the NH method. Panel~(\textbf{a}) demonstrates the convergence of the NOH method, while in (\textbf{b}), a comparison of the convergence rates between the NOH and the NH methods is conducted. For~both methods, we set $\delta M = 25$ in each iteration. For~NOH, $\epsilon$ was set to $0$, and~the iterative procedure was terminated when $\delta P \leq \epsilon$ for the first~time. (\textbf{a}) Convergence of the NOH~method. (\textbf{b}) Comparison between the NOH and NH~{methods.} 
	}
	\label{fig:Ex33}
\end{figure}

The convergence of the NOH method is illustrated  and compared with that of the NH method in \Cref{fig:Ex33}. 
{Although both methods converged quickly, NOH converged in 
	noticably fewer iterations compared to the NH method. With~only $100$ evaluations, the~NOH method could accurately estimate the failure probability as $P_f^h = 2.80 \times 10^{-4}$, with~a relative error of $3.70\%$, while the NH method only achieved $11.11\%$ while utilizing $875$~evaluations. The~faster convergence and lower relative error observed in the NOH method signify the precision and capabilities of the neural operator surrogate when compared to the neural~surrogate.

	These results highlight the potential of neural operator surrogates to significantly enhance computational efficiency and accuracy in a variety of applications.  Consequently, the~NOH method emerges as a more efficient and effective approach for estimating  failure probability.   }
\section{Conclusions}
\label{sec:conclusions}
This paper introduced a neural operator hybrid method for the estimation of  failure probability. Instead of approximating the limit state function directly, we reframe the problem as an operator learning task. This allows us to construct a highly efficient and precise surrogate operator model that can accurately estimate the limit state function. By~integrating the surrogate operator model into the hybrid algorithm, we created the neural operator hybrid method. The numerical results demonstrate that the proposed method provides an efficient strategy for estimating failure probability, particularly in systems governed by ODEs, multivariate functions, and~the Helmholtz equation. {Our proposed method exhibited superior performance  to that of the basic MCS approach, particularly in terms of efficiency. Furthermore, it surpassed the previous neural hybrid method in both efficiency and accuracy. Consequently, it is applicable and beneficial for addressing general failure probability estimation problems.} 
{The obtained results not only demonstrate the efficacy of neural operator learning frameworks in the context of failure probability estimation, but also imply their promising potential in other areas, such as Bayesian inverse problems and partial differential equations with random inputs.} In our future work, techniques such as importance sampling (IS) \cite{Xiu2011,IS2022} or adaptive learning~\cite{Lieu2022,Ada2021} can further reduce the  sample size required in the estimation process.

\bibliographystyle{spmpsci}       
\bibliography{mdpi-arxiv}                

\end{document}